\documentclass[conference]{IEEEtran}
\IEEEoverridecommandlockouts
\usepackage{cite}
\usepackage{amsmath,amssymb,amsfonts}
\usepackage{algorithmic}
\usepackage{graphicx}
\usepackage{subfigure}
\usepackage{textcomp}
\usepackage{xcolor}
\usepackage{stfloats}

\def\BibTeX{{\rm B\kern-.05em{\sc i\kern-.025em b}\kern-.08em
    T\kern-.1667em\lower.7ex\hbox{E}\kern-.125emX}}
\begin{document}
\title{A Novel 3D Non-Stationary Multi-Frequency Multi-Link Wideband MIMO Channel Model}
\DeclareRobustCommand*{\IEEEauthorrefmark}[1]{%
	\raisebox{0pt}[0pt][0pt]{\textsuperscript{\footnotesize\ensuremath{#1}}}}
\author{\IEEEauthorblockN{Li Zhang\IEEEauthorrefmark{1,2},
		Xinyue Chen\IEEEauthorrefmark{1,2}, Zihao Zhou\IEEEauthorrefmark{1,2}, Cheng-Xiang Wang\IEEEauthorrefmark{1,2*}, Chun Pan\IEEEauthorrefmark{3}, and
		Yungui Wang\IEEEauthorrefmark{3}}
	\IEEEauthorblockA{$^1$ National Mobile Communication Research Labratory, School of Information Science and Engineering,\\ Southeast University, Nanjing 210096, China \\ $^2$ Purple Mountain Labratories, Nanjing 211111, China\\$^3$ Huawei Technologies Co., Ltd., Nanjing 210012, China\\$^*$ Corresponding Author: Cheng-Xiang Wang\\
		Email: \{li-zhang,
		chenxinyue\_2019,
		zhouzh,
		chxwang\}@seu.edu.cn, \{panchun, wangyungui\}@huawei.com}}
\maketitle
\begin{abstract}
In this paper, a multi-frequency multi-link three-dimensional (3D) non-stationary wideband multiple-input multiple-output (MIMO)  channel model is proposed. The spatial consistency and multi-frequency correlation are considered in parameters initialization of every single-link and different frequencies, including large scale parameters (LSPs) and small scale parameters (SSPs). Moreover, SSPs are time-variant and updated when scatterers and the receiver (Rx) are moving. The temporal evolution of clusters is modeled by birth and death processes. The single-link channel model which has considered the inter-correlation can be easily extended to multi-link channel model. Statistical properties, including spatial cross-correlation function (CCF), power delay profile (PDP), and correlation matrix collinearity (CMC) are investigated and compared with the 3rd generation partner project (3GPP) TR 38.901 and quasi deterministic radio channel generator (QuaDRiGa) channel models. Besides, the CCF is validated against measurement data.
\end{abstract}
\section{Introduction}
\indent With the rapid development of wireless communication technology, spectrum resources are increasingly scarce\cite{b5} and user groups have higher requirements for the throughput and rate of communication system. An accurate channel model will be helpful to system evaluation. There exist phenomena that one application uses several frequency bands, i.e., WiFi working at 2.4 GHz and 5 GHz. 
%Recently, the WCR (World Radiocommunication Conference)-19 conference has decided to list the use of 6 GHz (6425-7125 MHz) in WRC-23 conference. It may be a potential new frequency band for wireless local area network (WLAN) system with larger bandwidth. 
Hence, frequency correlation needs to be explored and utilized.
% There exists literature that described how channel parameters change with carrier frequency. 
The frequency dependency of the delay spread was studied in\cite{multifreq1}.
3GPP TR38.901 (3GPP is used in the following text)\cite{b1} and QuaDRiGa\cite{b2} channel models proposed a multi-frequency correlation channel model which considers the frequency correlation.\\
\indent In terms of non-stationarity, many measurements have shown that the stationary interval is actually shorter than the simulation time in high mobility scenarios\cite{b6} . Channel models based on wide-sense assumption may ignore the characteristics of fast fading channel, such as the non-stationarity \cite{b7}. Reference\cite{highmobility1} demonstrated the necessity of establishing time-variant parameters in vehicle-to-vehicle scenarios to capture the channel non-stationarity. Reference\cite{highmobility2} drew the same conclusion in high-speed train scenarios.  Besides, long time evolution needs to take the birth and death of clusters into account. However, 3GPP does not consider the movement of scatterers and only Doppler frequency is a time-variant parameter. Although there is a birth-death process in QuaDRiGa channel model, the cluster positions keep fixed during a simulation segment. The Markov-process-based method of cluster birth and death\cite{{b7},{b9},{b10}} will lead to more realistic behaviors.\\
\indent Multi-link communications are important concepts to improve the realistic performance of communication systems\cite{multilink1}. However, most commonly used channel models such as WINNER\cite{b3} and ITU\cite{b8} channel models are drop-based, meaning that the scattering environment is randomly created for each link, contrast to the reality that mobile terminals close to each other will have similar parameters. Besides, when considering the movement of user ternimals or scatterers, parameters should be time-variant and experience smooth transition. Reference\cite{multilink1} characterized large-scale parameters in multi-link systems. Joint channel characteristics were modeled in a multi-link high-speed railway scenario\cite{multilink2}. Both 3GPP and QuaDRiGa channel models have considered the spatial consistency. In fact, the single-link channel models can easily be extended to multi-link channel models after considering the spatial correlation when initializing parameters. Therefore, it's of vital importance to include the spatial consistency into channel model.\\   
\indent In this paper, we proposed a multi-frequency multi-link 3D non-stationary wideband MIMO channel model, which compensates for the deficiency of 3GPP and QuaDRiGa channel models in terms of multi-frequency, multi-link and time-variant channel modelling. Detailedly we introduced the movement of the scatterers into QuaDRiGa channel model and consider the frequency correlation and spatial consistency, simultaneously.\\
\indent The remainder of this paper is organized as follows. In Section II, the proposed model is described in detail, which includes channel coefficient, initialization, drifting of parameters, and cluster evolution. Statistical properties such as spatial CCF, PDP, and CMC, are studied in Section III. In Section IV, simulation results of the proposed model and the other two standardized models are compared and analyzed. Conclusions are finally drawn in Section V.
\section{Description of the Channel Model}
\begin{figure}[htbp]
	\centering
	\includegraphics[width=0.45\textwidth]{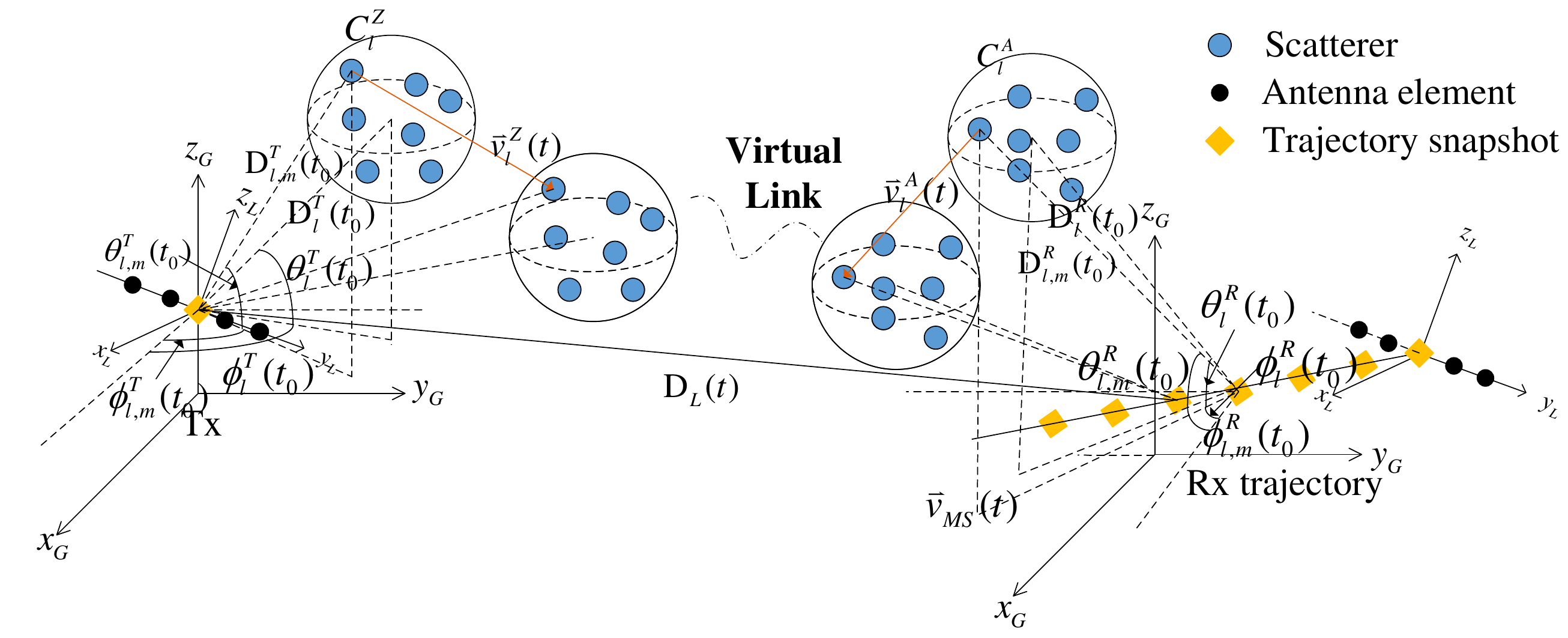}
	\caption{Non-stationary single-link indoor channel model.}
	\label{fig1}
\end{figure}
\begin{table*}
	\caption{Definition of Key Channel Model Parameters}
	\begin{center}
		\begin{tabular}{|c|c|}
		\hline
			\textbf{Symbol}&\textbf{Definition}\\
			\hline
			$C_l^Z, C_l^A$ & The first- and last-bounce clusters of $l$-th path, respectively\\
			\hline
			${\overset{\scriptscriptstyle\rightharpoonup}{v}}_l^Z(t)$, ${\overset{\scriptscriptstyle\rightharpoonup}{v}}_l^A(t)$, ${\overset{\scriptscriptstyle\rightharpoonup}{v}}_{MS}(t)$ & The velocities of the $l$-th first-bounce,  last-bounce cluster, and Rx,  respectively\\
			\hline
			$\phi_{l}^R(t)$, $\phi_{l}^T(t)$ & Azimuth angles of the $l$-th cluster at the Rx and Tx sides, respectively\\
			\hline
			$\phi_{l,m}^R(t)$, $\phi_{l,m}^T(t)$& Azimuth angles of the $m$-th ray of the $l$-th cluster at the Rx and Tx sides, respectively\\
			\hline
			$\phi_{L}^R(t)$, $\phi_{L}^T(t)$ & Azimuth angles of the LOS component at the Rx and Tx sides, respectively\\
			\hline
			$\theta_{l}^R(t)$, $\theta_{l}^T(t)$ & Elevation angles of the $l$-th cluster at the Rx and Tx sides, respectively\\
			\hline
			$\theta_{l,m}^R(t)$, $\theta_{l,m}^T(t)$  & Elevation angles of the $m$-th ray of the $l$-th cluster at the Rx and Tx sides, respectively\\
			\hline
			$\theta_{L}^R(t)$, $\theta_{L}^T(t)$ & Elevation angles of the LOS component at the Rx and Tx sides, respectively\\
			\hline
			$D_l^R(t) \left(D_l^T(t)\right)$ & Distance from the Rx (Tx) to $C_l^A  (C_l^Z)$ \\
			\hline
			$D_{l,m}^R(t_0) \left(D_{l,m}^T(t_0)\right)$ & Distance between the Rx (Tx) and $m$-th ray within $l$-th cluster \\
			\hline
			$D_{L}(t)$ & Distance between the Rx and the Tx \\
			\hline
%			$N(t)（$ & The number of clusters \\
%			\hline
		\end{tabular}
		\label{tab1}
	\end{center}
\end{table*}
\indent The proposed MIMO channel model is illustrated in Fig.~\ref{fig1}, with the transimitter (Tx) equipped with $M_T$ antennas and the Rx equipped with $M_R$ antennas. The antenna positions are given in local coordinate system. Considering a multi-bounce scattering propagation, only the $l$-th ($l=1,2,...,N(t)$) path is illustrated for clarity. $N(t)$ is the total number of paths in the link between the $s$-th $(s=1,2,...,M_T)$  transmitting antenna and the $u$-th $(u=1,2,...,M_R)$ receiving antenna at time $t$. $C_l^Z$ and $C_l^A$ denote the first-bounce cluster at Tx side and the last-bounce cluster at Rx side of $l$-th path, respectively. The propagation between $C_l^Z$ and $C_l^A$ is abstracted by a virtual link\cite{b3}, where signal will experience diffraction, reflection, and scattering, resulting in certain propagation delay. In the model, the Tx keeps static and the movement of the Rx as well as scatterers are described as different trajectories, i.e., linear track and circular track in global coordinate system. The velocities of  $C_l^Z$,  $C_l^A$, and the Rx are denoted by ${\overset{\scriptscriptstyle\rightharpoonup}{v}}_l^Z(t)$,  ${\overset{\scriptscriptstyle\rightharpoonup}{v}}_l^A(t)$, and ${\overset{\scriptscriptstyle\rightharpoonup}{v}}_{MS}(t)$ respectively, which can be calculated from the adjacent snapshots of their trajectories. Some other important parameters in the channel model are given in Table~\ref{tab1}.\\
%\indent The trajectory of the transceiver is generated according to the simulation environment.$\mathbf{P}^T(t)=(x^T(t),y^T(t),z^T(t))^T$ and $\mathbf{P}^R(t)=(x^R(t),y^R(t),z^R(t))^T$ are the positions of BS and MS in global coordinate system(GCS) at time $t$.$(\cdot)^T$ means the transpose operation of a vector. If considering the movement of clusters,their trajectory can be set as well.Similarly, $\mathbf{P}_l^Z(t)=(x_l^Z(t),y_l^Z(t),z_l^Z(t))^T$, $\mathbf{P}_l^A(t)=(x_l^A(t),y_l^A(t),z_l^A(t))^T$represent the positions of the first-bounce cluster and last-bounce cluster at time $t$.We can calculate $D_l^T(t_0)=||\mathbf{P}_l^Z(t_0) -\mathbf{P}^T(t_0)||_F$,where$||\cdot ||_F^2$ is Frobenius norm。$D_l^R(t_0)$ can be obtained by the same way.
\indent Considering small scale fading, path loss, and shadowing fading, the complete channel matrix of single-link is given by 
\begin{equation}
\label{1}
H={{[PL\cdot SF]}^{1/2}}{{H}_{s}}
\end{equation}
where $PL$ stands for the path loss and $SF$ stands for the shadowing fading. Besides, the complex matrix $H_s=[h_{u,s,f}(t,\tau )]_{M_R\times M_T}$ denotes small scale fading.
\subsection{Channel Impulse Response (CIR)}
The component $h_{u,s,f}(t,\tau)$ which denotes the CIR between the $u$-th receiving antenna and the $s$-th transmitting antenna at the $f$-th ($f=1,2,...,F$) center frequency is the superposition of the line-of-sight (LOS) component and non-line-of-sight (NLOS) components and can be expressed as
\begin{equation}
\label{2}
{{h}_{u,s,f}}(t,\tau )=h_{u,s,f}^{\text{NL}}(t,\tau )+h_{u,s,f}^{\text{L}}(t,\tau ).
\end{equation}
\indent The LOS component and NLOS components are shown at the top of the next page, where $(\cdot)^{*}$ means the transpose operation of a vector, $K(t)$ is the time-variant K-factor.
\begin{figure*}[ht]
%\begin{align}
%\label{3}
%H_{u,s,f}^{\text{L}}(t,\tau)=&\sqrt{\frac{K(t)}{K(t)+1}}{{\left[ \begin{matrix}
%		{{F}_{u,f,\theta }}\left( {{\theta }_{L}^R(t)},{{\phi }_{L}^R(t)} \right)  \\
%		{{F}_{u,f,\phi }}\left( {{\theta }_{L}^R(t)},{{\phi }_{L}^R(t)} \right)  \\
%		\end{matrix} \right]}^{*}}
%	\left[ \begin{matrix}
%\exp (j\Phi _{L}^{\theta\theta}) & 0  \\
%0 & -\exp (j\Phi _{L}^{\phi\phi})  \\
%\end{matrix} \right] \left[ \begin{matrix}
%{{F}_{s,f,\theta }}\left( {{\theta }_{L}^T(t)},{{\phi }_{L}^T(t)} \right)  \\
%{{F}_{s,f,\phi }}\left( {{\theta }_{L}^T(t)},{{\phi }_{L}^T(t)} \right)  \\
%\end{matrix} \right] \\ &\times\exp \left( j(\psi_{u,s,L}(t)+\psi_{L}^0) \right)\delta(\tau-\tau_{L}(t)) \nonumber 
%\end{align}
\begin{align}
\label{3}
H_{u,s,f}^{\text{L}}(t,\tau)=&\sqrt{\frac{K(t)}{K(t)+1}}{{\left[ \begin{matrix}
		{{F}_{u,f,\theta }}\left( {{\theta }_{L}^R(t)},{{\phi }_{L}^R(t)} \right)  \\
		{{F}_{u,f,\phi }}\left( {{\theta }_{L}^R(t)},{{\phi }_{L}^R(t)} \right)  \\
		\end{matrix} \right]}^{*}}
\left[ \begin{matrix}
1 & 0  \\
0 & -1  \\
\end{matrix} \right] \left[ \begin{matrix}
{{F}_{s,f,\theta }}\left( {{\theta }_{L}^T(t)},{{\phi }_{L}^T(t)} \right)  \\
{{F}_{s,f,\phi }}\left( {{\theta }_{L}^T(t)},{{\phi }_{L}^T(t)} \right)  \\
\end{matrix} \right] \\ &\times\exp \left( j(\psi_{u,s,L}(t)+\psi_{L}^0) \right)\delta(\tau-\tau_{L}(t)) \nonumber 
\end{align}
\begin{align}
\label{4}
H_{u,s,f}^{\text{NL}}(t,\tau)=&\sqrt{\frac{1}{K(t)+1}}\sum\limits_{l=1}^{N(t)}\sum\limits_{m=1}^{M}\sqrt{\frac{{{P_{l,f}}(t)}}{M}}
\begin{bmatrix}
{{F}_{u,f,\theta }}\left( {{\theta }_{l,m}^R(t)},{{\phi }_{l,m}^R(t)} \right)  \\
{{F}_{u,f,\phi }}\left( {{\theta }_{l,m}^R(t)},{{\phi }_{l,m}^R(t)} \right)  \\
\end{bmatrix}  ^{*}  \begin{bmatrix}
\exp \left( j\Phi _{l,m}^{\theta \theta } \right) & \sqrt{\frac{1}{{\kappa }_{l,m}(t)}}\exp \left( j\Phi _{l,m}^{\theta \phi } \right)  \\
\sqrt{\frac{1}{{\kappa }_{l,m}(t)}}\exp \left( j\Phi _{l,m}^{\phi \theta } \right) & \exp \left( j\Phi _{l,m}^{\phi \phi } \right)  \\
\end{bmatrix} \\
&\times \begin{bmatrix}
{{F}_{s,f,\theta }}\left( {{\theta }_{l,m}^T(t)},{{\phi }_{l,m}^T(t)} \right)  \\
{{F}_{s,f,\phi }}\left( {{\theta }_{l,m}^T(t)},{{\phi }_{l,m}^T(t)} \right)  \\
\end{bmatrix}\nonumber 
\exp \left( j(\psi_{u,s,l,m}(t)+\psi_{l,m}^0) \right)\delta(\tau-\tau_l(t))\nonumber 
\end{align}
\hrulefill
\end{figure*}
%The $\Phi _{L}^{\theta\theta}$ and $\Phi _{L}^{\phi\phi}$ are the random intial phases of the LOS path for $\left\{\theta\theta,\phi\phi\right\}$ polarization combinations, respectively.
For the LOS component (\ref{3}), $F_{u,f,\theta} (F_{s,f,\theta})$ and $F_{u,f,\phi} (F_{s,f,\phi})$ are the radiation patterns of antenna element $u(s)$ at the $f$-th frequency for vertical and horizontal polarizations, respectively. Besides, the phase caused by propagation distance can be represented by
\begin{equation}
\label{5}
\psi_{u,s,L}(t)=\frac{2\pi}{\lambda}\cdot(d_{u,s,L}(t))
%\psi_{u,s,L}(t)={2\pi}/{\lambda}\cdot(d_{u,s,L}(t) \text{mod} \lambda)
\end{equation} 
which contains the Doppler effect, where 
\begin{equation}
\label{6} d_{u,s,L}(t)=\hat{r}_{rx,L}^\mathrm{*}(t)\cdot{{{\overset{\scriptscriptstyle\rightharpoonup}{d}}}_{rx,u}}+\hat{r}_{tx,L}^\mathrm{*}(t)\cdot{{{\overset{\scriptscriptstyle\rightharpoonup}{d}}}_{tx,s}}+d_{1,1,L}(t)
\end{equation}
where $d_{1,1,L}(t)$ is the distance between the first transmitting antenna and receiving antenna. The spatially correlated random initial phase of LOS path is denoted by $\psi_{L}^0$. Moreover, the delay of LOS path is given by
\begin{equation}
\label{7}
\tau_{L}(t)=D_{L}(t)/c
\end{equation}
where $c$ represents the speed of light. For the NLOS componnets (\ref{4}), $\hat{r}_{rx,l,m}(t)$ and ${\hat{r}}_{tx,l,m}(t)$ are the spherical unit vectors with arrival angles and departure angles, respectively. $\psi_{l,m}^0$ is the spatially correlated random initial phase of each ray. ${\kappa }_{l,m}(t)$ is the cross-polarization ratio (XPR).\\
\indent Similarly, the phase caused by the distance between the $s$-th transmitting antenna and the $u$-th receiving antenna via the $m$-th ray within the $l$-th cluster can be calculated by
\begin{equation}
\label{8}
\psi_{u,s,l,m}(t)=\frac{2\pi}{\lambda}(d_{u,s,l,m}(t))
%\psi_{u,s,l,m}(t)={2\pi}/{\lambda}(d_{u,s,l,m}(t) \text{mod} \lambda)
\end{equation}
which also includes the Doppler effect, where
\begin{equation}
\label{9}
d_{u,s,l,m}(t)=\hat{r}_{rx,l,m}^{*}.{{{\overset{\scriptscriptstyle\rightharpoonup}{d}}}_{rx,u}}+\hat{r}_{tx,l,m}^{*}.{{{\overset{\scriptscriptstyle\rightharpoonup}{d}}}_{tx,s}}+d_{1,1,l,m}(t)
\end{equation}
where $d_{1,1,l,m}(t)$ is the distance traveling from the first transmitting antenna via the $m$-th ray within the $l$-th cluster to the first receiving antenna. It can be calculated by 
\begin{equation}
\label{10}
d_{1,1,l,m}(t)=D_{1,l,m}^T(t)+D_{1,l,m}^R(t)+D_{l,m}^{ZA}(t)
\end{equation} 
where $D_{1,l,m}^T(t)$ and $D_{1,l,m}^R(t)$ are the distance of the first antenna to the $m$-th ray at the Tx and Rx sides, respectively. The $D_{l,m}^{ZA}(t)$ is the distance calculated by the cluster positions.\\
%\indent $\hat{r}_{rx,l,m}(t)$ and ${\hat{r}}_{tx,l,m}(t)$ are the spherical unit vectors with arrival angles and departure angles, respectively.
% \begin{equation}
%\tag{11}
%\label{11}
%{{\hat{r}}_{rx,l,m}}(t)=\left[ \begin{matrix}
%\cos {{\theta }_{l,m,EOA}}(t)\cos {{\phi }_{l,m,AOA}}(t)  \\
%\cos {{\theta }_{l,m,EOA}}(t)\sin {{\phi }_{l,m,AOA}}(t)  \\
%\sin {{\theta }_{l,m,EOA}}(t)  \\
%\end{matrix} \right].
%\end{equation}
%\indent ${\hat{r}}_{tx,l,m}(t)$ is the spherical unit vector with azimuth departure angle ${\phi }_{l,m,AOD}$ and elevation departure angle ${\theta }_{l,m,EOD}$, given by
%\begin{equation}
%\tag{12}
%\label{12}
%{{\hat{r}}_{tx,l,m}}(t)=\left[ \begin{matrix}
%\cos {{\theta }_{l,m,EOD}}(t)\cos {{\phi }_{l,m,AOD}}(t)  \\
%\cos {{\theta }_{l,m,EOD}}(t)\sin {{\phi }_{l,m,AOD}}(t)  \\
%\sin {{\theta }_{l,m,EOD}}(t)  \\
%\end{matrix} \right].
%\end{equation}
\indent According to the new method in \cite{b2}, to ensure the given cluster powers are  the same to those calculated by channel coefficients,  the channel amplitude of all snapshots needs to be scaled. Just to mention that the time interval between adjacent snapshots is time $\Delta t$ in this paper. Let $h_{u,s,f,l,m,p}^{\text{NL}}$ denote the CIR of the $p$-th snapshot, the $m$-th ray within the $l$-th cluster at the $f$-th center frequency, the CIR of the $p$-th snapshot, the $l$-th cluster is $h_{u,s,f,l,p}^{\text{NL}}=\sum\limits_{m=1}^{M}h_{u,s,f,l,m,p}^{\text{NL}}$. The scaled channel coefficient is given by
\begin{equation}
%\tag{13}
\label{11}
\tilde{h}_{u,s,f,l,p}^{\text{NL}}=\sqrt{\frac{{{P}_{l,f}}}{M}\cdot \frac{\sum\limits_{p=1}^{P}{\sum\limits_{m=1}^{M}{{{\left| h_{u,s,f,l,m,p}^{\text{NL}} \right|}^{2}}}}}{\sum\limits_{p=1}^{P}{{{\left| h_{u,s,f,l,p}^{\text{NL}} \right|}^{2}}}}}\cdot h_{u,s,f,l,p}^{\text{NL}}.
\end{equation}
\subsection{Correlated LSPs}\label{AA}
Together, there are eight LSPs, including delay spread (DS), K-factor, $SF$, elevation spread of departure (ESD), elevation spread of arrival (ESA), azimuth spread of departure (ASD), azimuth spread of arrival (ASA), and XPR. To ensure the continuity of parameter variation, the sum of sinousoids (SoS) method in \cite{b2} is adopted to generate the LSPs. The distribution parameters of the LSPs are derived from \cite{b1}. \\
\indent After done with all LSPs, the large-scale fading of the channel in the linear domain can be obtained by multiplying the shadowing fading and path loss :\\
\begin{equation}
%\tag{14}
\label{12}
PL\cdot SF=\sqrt{{{10}^{0.1\left( P{{L}^{[dB]}}+S{{F}^{[dB]}} \right)}}}.
\end{equation}
\indent For long time evolution where the traveling distance the distance is beyond correlation distance, the channel modeling needs to take into account the update of LSPs, which will have an impact on the cluster power, and the update of the cluster delay and angle is accomplished in the drifting process.
%\indent As known from Table 7.5-6 of the \cite{b1}, the correlation distance of LSPs is within 10 m whether it is the LOS scenario or the NLOS scenario. Considering the longest moving trajectory of indoor office can be around 25 m, the channel modeling needs to take into account the update of LSPs, which will have an impact on the cluster power, and the update of the cluster delay and angle is accomplished in the drifting process.

\subsection{Multi-Frequency SSPs}\label{AA}
It cannot be arbitrary to initialize the SSPs of the cluster, since these parameters are correlated to  some extent at two positions nearby. Therefore, SoS method is also adopted to initialize the cluster angles and delays satisfying spatial consistency.\\
\indent Firstly, according to the initial position vectors $\mathbf{P}^{T}(t_0)$ and $\mathbf{P}^{R}(t_0)$, the spatially correlated uniform distributed azimuth angles ${{\tilde{\phi }}}_{l}^T$, ${{\tilde{\phi }}}_{l}^R$, and elevation angles ${{\tilde{\theta }}}_{l}^T$, ${{\tilde{\theta }}}_{l}^R$ can be obtained by \eqref{13}, and the initial delay is shown in ~\eqref{14}.
% \begin{equation}
%\tag{15}
%\label{15}
%\begin{split}
%& {{{\tilde{\phi }}}_{l,AOD/AOA}}\left( {{\mathbf{P}}^{T}}({{t}_{0}}),{{\mathbf{P}}^{R}}({{t}_{0}}) \right)\\&=\frac{\pi }{2}erfc\left( \frac{\widetilde{X}_{l}^{\phi d}\left( {{\mathbf{P}}^{T}}({{t}_{0}}) \right)+\widetilde{X}_{l}^{\phi a}\left( {{\mathbf{P}}^{R}}({{t}_{0}}) \right)}{2} \right)-\frac{\pi }{2} \\ 
%\end{split}
%\end{equation}
\begin{equation}
%\tag{16}
\label{13}
\begin{split}
& {{{\tilde{\theta}}/{\tilde{\phi}}}}\left( {{\mathbf{P}}^{T}}({{t}_{0}}),{{\mathbf{P}}^{R}}({{t}_{0}}) \right)\\&=\frac{\pi }{2}erfc\left( \frac{\widetilde{X}_{l}^{(\theta/\phi) d}\left( {{\mathbf{P}}^{T}}({{t}_{0}}) \right)+\widetilde{X}_{l}^{(\theta/\phi) a}\left( {{\mathbf{P}}^{R}}({{t}_{0}}) \right)}{2} \right)-\frac{\pi }{2} \\ 
\end{split}
\end{equation}
  \begin{equation}
%\tag{17}
\label{14}
{{\tilde{\tau }}_{l}}(t_0)=-\ln \left\{ X_{l}^{\tau }\left( {{\mathbf{P}}^{T}}(t_0),{{\mathbf{P}}^{R}}(t_0) \right) \right\}
\end{equation}
where $\widetilde{X}_{l}^{\phi d}$, $\widetilde{X}_{l}^{\phi a}$, $\widetilde{X}_{l}^{\theta d}$, and $\widetilde{X}_{l}^{\theta a}$ are spatially dependent random variables that follow a normal distribution with zero-mean and unit variance. The $erfc(\cdot)$ is complementary error function. $X_{l}^{\tau }$ is a spatially correlated uniformly distributed random variable ranging from 0 to 1.
%\begin{equation}
%\tag{21}
%\label{21}
%\begin{split}
%&X_{l}^{\tau}\left( \mathbf{P}_{l}^{Z}(t_0),\mathbf{P}_{l}^{A}(t_0) \right)\\ &=\frac{1}{2}erfc\left( \frac{\widetilde{X}_{l}^{\tau }\left( \mathbf{P}_{l}^{Z}(t_0) \right)+\widetilde{X}_{l}^{\tau }\left( \mathbf{P}_{l}^{A}(t_0) \right)}{2\sqrt{{{\rho }_{\tau }}\left( {{d}_{tr}} \right)+1}} \right)\\
%\end{split}
%\end{equation}
The initial delay $\tilde{\tau}_l$ and angles ${\tilde{\phi}}_{l}^T$, ${\tilde{\phi}}_{l}^R$, ${\tilde{\theta}}_{l}^T$, ${\tilde{\theta}}_{l}^R$ are assumed to be frequency-independent. However, DS and angular spread (AS) are generally frequency-dependent. Thus different cluster powers can be achieved under different frequencies. The initial power is calculated by \\
\begin{equation}
%\tag{18}
\label{15}
\begin{split}
{{{\tilde{P}}}_{l,f}}({{t}_{0}})=&\exp \left\{ -{\widetilde{\tau }_{l}}({{t}_{0}})\cdot g_{f}^{DS}-{{\left( {{{\tilde{\phi }}}_{l}^T}({{t}_{0}}) \right)}^{2}}\cdot g_{f}^{ASD}\right. \\ 
& \left.-{{\left( {{{\tilde{\phi }}}_{l}^R}({{t}_{0}}) \right)}^{2}}\cdot g_{f}^{ASA}  -\left| {{{\tilde{\theta }}}_{l}^T}({{t}_{0}}) \right|\cdot g_{f}^{ESD}\right. \\ 
& \left.-\left| {{{\tilde{\theta }}}_{l}^R}({{t}_{0}}) \right|\cdot g_{f}^{ESA} \right\}   \\ 
\end{split}
\end{equation}

\begin{equation}
%\tag{19}
\label{16}
%{{P}_{l,f}}({{t}_{0}})=\frac{{{{\tilde{P}}}_{l,f}}({{t}_{0}})}{\sum\limits_{l=1}^{N({{t}_{0}})}{{{{\tilde{P}}}_{l,f}}({{t}_{0}})}}
{{P}_{l,f}}({{t}_{0}})={{{\tilde{P}}}_{l,f}}({{t}_{0}})/\sum\limits_{l=1}^{N({{t}_{0}})}{{{{\tilde{P}}}_{l,f}}({{t}_{0}})}
\end{equation}
where $g_{f}^{DS}$, $g_{f}^{ASD}$, $g_{f}^{ASA}$, $g_{f}^{ESD}$, and $g_{f}^{ESA}$ are constant for single frequency but exist frequency-dependency if there are multiple frequencies. The calculation method is seen in \cite{b2}. In order to ensure the AS and DS calculated by the power and angle is consistent with the value generated during LSP initialization, the initial angles and delays needs to be scaled and adjusted. The detailed information can be referred to \cite{b2}.\\
\indent Finally, we need to consider the effect of the LOS angle that allow the construction of two rotation matrices in Cartesian coordinates, one for the Tx and one for the Rx. The final angles are derived from $\hat{\mathbf{c}}_{l}$ that is obtained by applying rotation matrix.\\
%The NLOS departure and arrival angles from previous calculation are converted to Cartesian coordinate first to obtain  $\mathbf{c}_{l}$. The $\hat{\mathbf{c}}_{l}$ is obtained by applying rotation matrix that is constructed by LOS angles. The final azimuth and elevation angles are obtained by converting $\hat{\mathbf{c}}_{l}$ to spherical coordinates.\\
%\begin{equation}
%\tag{35}
%\label{35}
%\mathbf{c}_{l}=\left[\begin{array}{c}
%\cos \theta_{l} \cdot \cos \phi_{l} \\
%\cos \theta_{l} \cdot \sin \phi_{l} \\
%\sin \theta_{l}
%\end{array}\right].
%\end{equation} 
%\begin{equation}
%\tag{36}
%\label{36}
%\left[\begin{array}{ccc}
%\cos \theta_{L O S} \cdot \cos \phi_{L O S} & -\sin \phi_{L O S} & \sin \theta_{L O S} \cdot \cos \phi_{L O S} \\
%\cos \theta_{L O S} \cdot \sin \phi_{L O S} & \cos \phi_{L O S} & \sin \theta_{L O S} \cdot \sin \phi_{L O S} \\
%\sin \theta_{L O S} & 0 & \cos \theta_{L O S}
%\end{array}\right]
%\end{equation}
\begin{equation}
%\tag{22}
\label{17}
\begin{array}{l}
\phi_{l}^{R/T}(t_0)=\arctan _{2}\left\{\hat{c}_{1, x}, \hat{c}_{1, y}\right\} \\
\theta_{l}^{R/T}(t_0)=\arctan _{2}\left\{\hat{c}_{l, z}, \sqrt{\hat{c}_{l, x}^{2}+\hat{c}_{l, y}^{2}}\right\}
\end{array}
\end{equation}
where $\mathrm{arctan}_2\left\{\cdot\right\}$ is  the multi-valued inverse tangent. Finally apply the scenario-dependent cluster wise root mean square (RMS) AS $c_\sigma (\sigma=\text{ASA, ASD, ESA, ESD})$ and add offset angles $\alpha_{m}$ from Table 7.5-3 in \cite{b1} to the cluster angles
\begin{equation}
%\tag{23}
\label{18}
{{\phi }_{l,m}^{R/T}(t_{0})}={{\phi }_{l}^{R/T}(t_{0})}+\frac{\pi \cdot {{c}_{\sigma}}\cdot {\alpha}_{m}}{{180}^{o}}.
\end{equation}
%\begin{equation}
%\tag{38}
%\label{38}
%{{\phi }_{l.m,AOD}(t_{0})}={{\phi }_{l,AOD}(t_{0})}+\frac{\pi \cdot {{c}_{ASD}}\cdot {{{\hat{\phi }}}_{m}}}{{{180}^{o}}}
%\end{equation}
%\begin{equation}
%\tag{39}
%\label{39}
%{{\phi }_{l.m,AOA}(t_{0})}={{\phi }_{l,AOA}(t_{0})}+\frac{\pi \cdot {{c}_{ASA}}\cdot {{{\hat{\phi }}}_{m}}}{{{180}^{o}}}
%\end{equation}
%\begin{equation}
%\tag{40}
%\label{40}
%{{\theta }_{l.m,EOD}(t_{0})}={{\theta }_{l,EOD}(t_{0})}+\frac{\pi \cdot {{c}_{ESD}}\cdot {{{\hat{\phi }}}_{m}}}{{{180}^{o}}} 
%\end{equation}
%\begin{equation}
%\tag{41}
%\label{41}
%{{\theta }_{l.m,EOA}(t_{0})}={{\theta }_{l,EOA}(t_{0})}+\frac{\pi \cdot {{c}_{ESA}}\cdot {{{\hat{\phi }}}_{m}}}{{{180}^{o}}}
%\end{equation}
%\begin{table}[htbp]
%	\setlength{\abovecaptionskip}{0cm}
%	\setlength{\belowcaptionskip}{-0cm}
%	\centering
%%	\renewcommand{\captionlabelfont}{\small}
%	\caption{\small{Ray offset angles within a cluster}}
%	\begin{tabular}{|c|c|}
%		\hline
%		Ray number $m$ & Offset angles($^\circ$)\\
%		\hline
%		1, 2	& $\pm$0.0447 \\
%		\hline
%		3, 4	& $\pm$0.141 3\\
%		\hline
%		5, 6	& $\pm$0.2492\\
%		\hline
%		7, 8	& $\pm$0.3715\\
%		\hline
%		9, 10	& $\pm$0.5129\\
%		\hline
%		11, 12	& $\pm$0.6798\\
%		\hline
%		13, 14	& $\pm$0.8844\\
%		\hline
%		15, 16	& $\pm$1.1481\\
%		\hline
%		17, 18  & $\pm$1.5195\\
%		\hline
%		19, 20  & $\pm$2.1551\\
%		\hline		
%	\end{tabular}
%	\label{subpath}
%\end{table}\\
\indent The delay and subpath angles are then used to determine the initial positions of the first-bounce and the last-bounce clusters according the method in \cite{b2}.
\subsection{Time Evolution}\label{AA}
We introduced two characteristics to embody the non-stationarity of the proposed channel model\cite{b7}, namely the constant update of channel parameters and the cluster birth-death process. Therefore, there are two updating intervals in the proposed model, i.e., the sampling interval for channel coefficients $\Delta t$ and sampling interval for cluster birth-death process $\Delta t_{BD}$ which is identical to the update interval of LSPs. \\
\indent The existence of both moving Rx and moving clusters can result in the time variance of a wireless channel. Hence the variable $q(t+\Delta t_{BD})$ is introduced to measure how fast the propagation environment changes during the time interval $\Delta t_{BD}$ and can be used as a measure of the channel fluctuation, shown as $q(t+\Delta t_{BD})=q_{c}(t+\Delta t_{BD})+q_{r}(t+\Delta t_{BD})$, where $q_{r}(t+\Delta t_{BD})$ is the channel variance in time, resulting from the movement of the Rx, defined as $q_{r}(t+\Delta t_{BD})=\left\|\vec{v}_{M S}\right\| \Delta t_{BD}$. Similarly, the channel flucation caused by the moving scatterers is defined as $ q_{c}(t+\Delta t_{BD})=P_{c}\left(\left\|\vec{v}^{A}\right\|+\left\|\vec{v}^{Z}\right\|\right) \Delta t_{BD} $, where $P_{c}$ is the propability of cluster movements \cite{b9}. For simplicity, mean cluster velocities, i.e.,
$\vec{v}^{A}=E\left[\vec{v}_{l}^{A}\right], \vec{v}^{Z}=E\left[\vec{v}_{l}^{Z}\right]$ are used to calculate the probabilities of clusters at $t+t_{BD}$ survived from $t$.
\begin{equation}
%\tag{24}
\label{19}
{{P}_{surv}}(q(t+\Delta t_{BD}))={{e}^{-\frac{{{\lambda }_{R}}q(t+\Delta t_{BD})}{D_{c}^{a}}}}
\end{equation}
where $\lambda_{R}$ denotes the recombination rate of clusters. $D_c^a$ is the scenario depedent correlation factor. The newly generated clusters during the interval $\Delta t_{BD}$ is represented as:
\begin{equation}
%\tag{25}
\label{20}
E[{{N}_{ng}}(\Delta t_{BD})]=\frac{{{\lambda }_{G}}}{{{\lambda }_{R}}}(1-{{P}_{surv}})
\end{equation}
where $\lambda_G$ is the generation rate of clusters.
\subsection{Drifting of SSPs}\label{AA}
After the SSPs are initialized, their values are updated when the Rx and clusters move along trajectories, ensuring the spatial consistency for mobile terminals. Hence, the parameters and channel coefficients are updated at so-called snapshot positions. The distance vectors between the Tx and $m$-th ray within $C_l^Z$ is calculated by
\begin{equation}
%\tag{26}
\label{21}
\mathbf{D}_{l,m}^{T}(t)=\mathbf{D}_{l,m}^{T}({{t}-\Delta t})\text{+}\overset{\scriptscriptstyle\rightharpoonup}{v}_{l}^{Z}({{t}})\Delta t.
\end{equation}
\indent The azimuth and elevation departure angles can be determined by \eqref{22} and \eqref{23}. The operations of the Rx side are indentical.
\begin{equation}
%\tag{27}
\label{22}
{{\phi }_{l,m}^T}(t)={{\arctan }_{2}}\left\{ \mathbf{D}_{l,m}^{T}(t)\cdot \hat{y},\mathbf{D}_{l,m}^{T}(t)\cdot \hat{x} \right\}
\end{equation}
\begin{equation}
%\tag{28}
\label{23}
{{\theta }_{l,m}^T}(t)\text{=}\arcsin \left( \mathbf{D}_{l,m}^{T}(t)\cdot \hat{z} \right)
\end{equation}
\indent Based on the updated distance, the delay  at time $t$ can be achieved through \eqref{24}. The cluster powers can be obtained by substituting the updated cluster angles, delays, and LSPs into formula \eqref{15} and \eqref{16}. Note that the update interval is the same to that of LSPs and cluster birth and death.
\begin{equation}
%\tag{29}
\label{24}
\tau_{l}(t)=\tilde{\tau}_{l,link}(t)+\sum_{m=1}^M \frac{D_{l, m}^{T}(t)+D_{l, m}^{R}(t)+D_{l, m}^{Z A}(t)}{c \cdot M} 
\end{equation}
where $\tilde{\tau}_{l,link}(t)$ is the virtual delay between $C_l^Z$ and $C_l^A$, which is modeled as non-negative random variable with exponential distribution and generated by SoS method.

\section{Statistical Properties of the Proposed Model}
\subsection{Time-Variant PDP}\label{AA}
\indent The time-variant PDP $\Lambda(t, \tau)$ reveals the relationship between power and delay of multipaths, which is expressed as
\begin{equation}
%\tag{30}
\label{25}
\Lambda(t, \tau)=\sum_{l=1}^{N\left(t\right)} P_{l}(t) \delta\left(\tau-\tau_{l}(t)\right).
\end{equation}
\subsection{Time-Variant Transfer Function}\label{AA}
Consider the single frequency and single-link of the model, the time-variant transfer function is defined as the Fourier transform of time-variant CIR $h_{u,s}(t,\tau)$ with respect to time delay $\tau$ and can be expressed as
\begin{equation}
%\tag{31}
\label{26}
H_{u,s}(t, f)=\int_{-\infty}^{\infty} h_{u,s}(t, \tau) e^{-j 2 \pi f \tau} d \tau.
\end{equation}
\subsection{Local Space-Time-Frequency (STF) Correlation Function}\label{AA}
The local STF correlation function is defined as
\begin{equation}
%\tag{32}
\label{27}
\begin{split}
	&R_{u s, l^{\prime} s^{\prime}}\left(t, f ; \Delta t, \Delta f, \Delta d_{u}, \Delta d_{s}\right)\\ 
	 &=E\left\{H_{u, s}(t+\Delta t, f+\Delta f) H_{u^{\prime}, s^{\prime}}^{*}(t, f)\right\}.
\end{split}
\end{equation}
\indent Substituting transfer function into \eqref{27}, the STF correlation function can be exressed as the superposition of LOS and NLOS components, given as \eqref{28}.
\begin{equation}
%\tag{33}
\label{28}
%\begin{split}
\begin{array}{l}
R_{u s, u^{\prime} s^{\prime}}\left(t, f ; \Delta t, \Delta f, \Delta d_{u}, \Delta d_{s}\right) 
\\=\sqrt{\frac{K(t)}{K(t)+1}\frac{K(t+\Delta t)}{K(t+\Delta t)+1}} R_{u s, u^{\prime} s^{\prime}}^{\operatorname{L}}\left(t, f ; \Delta t, \Delta f, \Delta d_{u}, \Delta d_{s}\right)\\+\sqrt{\frac{1}{K(t)+1}\frac{1}{K(t+\Delta t)+1}} \sum\limits_{l=1}^{N\left(t\right)} R_{u s, u^{\prime} s^{\prime},l}^{\operatorname{NL}}\left(t, f ; \Delta t, \Delta f, \Delta d_{u}, \Delta d_{s}\right)
\end{array}
%\end{split}
\end{equation}
\indent  Let $\Delta f=0$, $\Delta d_{u}=0$ $(u=u^{\prime})$, and $\Delta d_{s}=0$ $(s=s^{\prime})$, temporal autocorrelation function (ACF) can be obtained. What's more, let $\Delta f=0, \Delta t=0$, the STF can be simplified to spatial CCF.
\subsection{CMC}\label{AA}
In the multi-link channel model, the correlation between two links is described by CMC which calculated by \eqref{31}. The value of CMC ranges from 0 to 1. The closer the CMC is to 1, the higher the mutual correlation of the two links is\cite{CMC}.
\begin{equation}
%\tag{36}
\label{31}
\text{CMC}=\frac{\left|t r\left\{\mathbf{R}_{1} \mathbf{R}_{2}^\mathrm{H}\right\}\right|}{\left\|\mathbf{R}_{1}\right\|_\mathrm{F}\left\|\mathbf{R}_{2}\right\|_\mathrm{F}}
\end{equation}
where $tr\left\{\cdot\right\}$ is trace of a matrix,  $\left\{\cdot\right\}^\text{H}$ is the conjugate transpose of a complex matrix, and $\left\|\cdot\right\|_\mathrm{F}$ is the Frobenius norm. $\mathbf{R}_i,i\in \left\{1,2\right\}$ is the correlation matrix of link $i$, shown as
\begin{equation}
%\tag{37}
\label{32}
%\begin{array}{c}
\mathbf{R}_{i}=\sum_{S=1}^{N_{S}}\left(\mathbf{H}_{i}(S)\right)^\mathrm{H} \mathbf{H}_{i}(S)
%\end{array}
\end{equation}
%\begin{equation}
%\tag{57}
%\label{57}
%%\begin{array}{c}
%R_{1}=\sum_{s=1}^{N_{s}}\left(H_{1, f}(t, f, s)\right)^\mathrm{H} H_{1, f}(t, f, s)
%%\end{array}
%\end{equation}
%\begin{equation}
%\tag{58}
%\label{58}
%%\begin{array}{c}
%R_{2}=\sum_{s=1}^{N_{s}}\left(H_{2, f}(t+\Delta t, f+\Delta f, s)\right)^{H} H_{2, f}(t+\Delta t, f+\Delta f, s)
%%\end{array}
%\end{equation}
where $\mathbf{H}_{i}(S)$ is the channel transfer matrix of $S$-th realization of $i$-th link at $f$-th frequency. $N_s$ is the total number of channel realizations. 
\section{Results and Analysis}
\begin{figure}[htbp]
	\centering
	\includegraphics[scale=0.15]{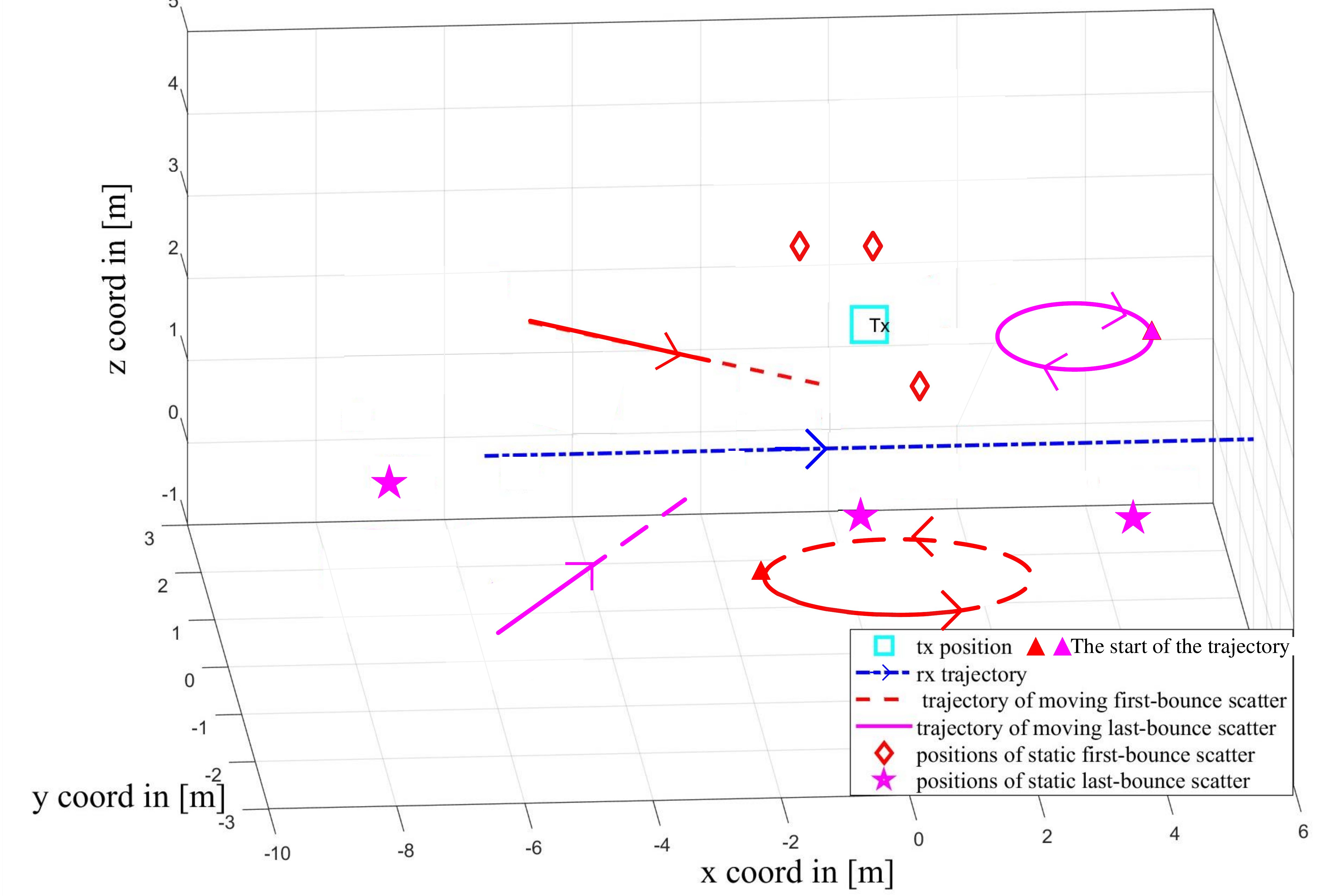}
	\caption{The visualization of trajectories of the Rx and clusters.}
	\label{trajectory}
\end{figure}

\begin{figure}[htbp]
	\subfigure[3GPP]{
	\begin{minipage}{2.85cm}
		\centering
		\includegraphics[width=2.85cm]{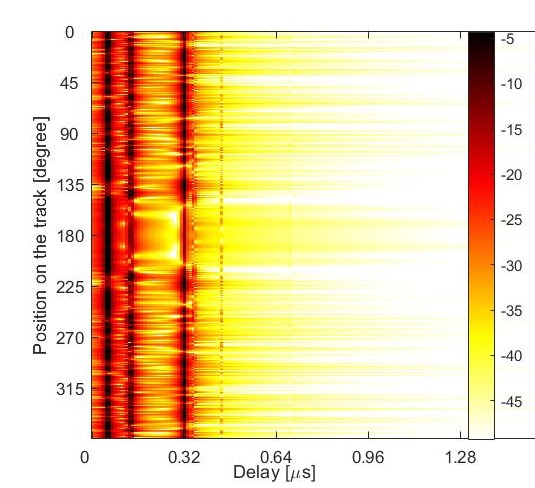}
%		\label{a}
\end{minipage}}
\subfigure[QuaDRiGa]{\begin{minipage}{2.85cm}
		\centering
		\includegraphics[width=2.85cm]{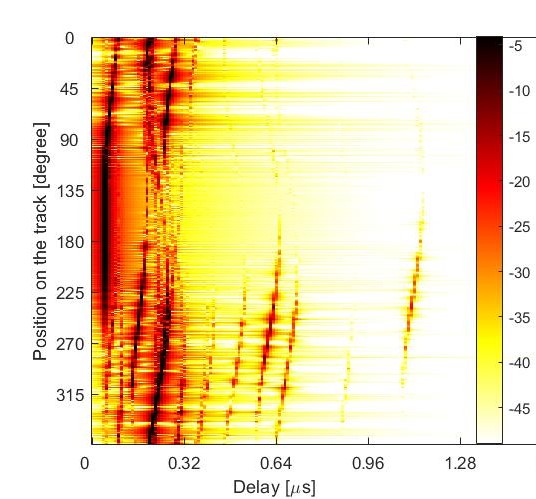}
%		\label{b}
\end{minipage}}
\subfigure[the proposed model]{\begin{minipage}{2.85cm}
		\centering
		\includegraphics[width=2.85cm]{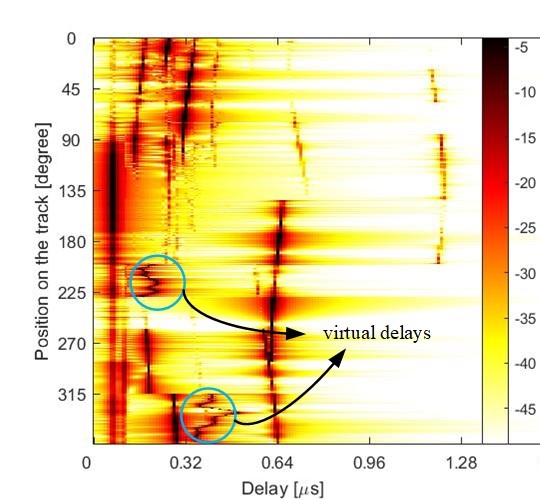}
%		\label{c}
\end{minipage}}
\caption{PDP of 3GPP, QuaDRiGa, and the proposed channel model ($N = 20$, ${\protect\overset{\scriptscriptstyle\rightharpoonup}{v}}_{MS}(t) = 3$ \text{m/s}, track length = 12 m, the simulation scenario is the simulated scenario is indoor office LOS scenario in\cite{b1}).}
\label{pdp}
\end{figure}

\begin{figure}[htbp]
  \centering
  	\includegraphics[width=8cm,height=6cm]{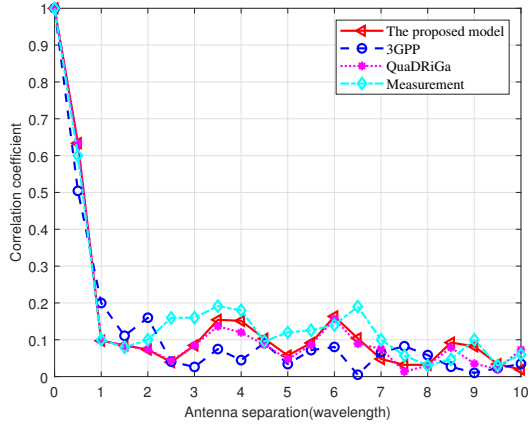}
   \caption{Spatial CCFs of 3GPP, QuaDRiGa, the proposed channel models, and measurement data in \cite{CCF_mea} ($f_c=5.2$ \text{GHz}, $M_T=8$, $M_R=141$, $\mu_{ASD}=0.98$, $\mu_{ESD}=1$, $c_{ASD}=c_{ESD}=1$, other parameters can be referred to\cite{b1} indoor office NLOS scenario).}
   \label{ccf_mea}
\end{figure}

\begin{figure}[htbp]
  \centering
  	\includegraphics[width=8cm,height=6cm]{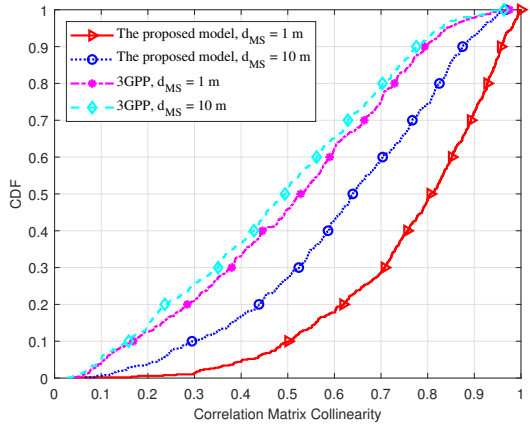}
   \caption{CDF of correlation matrix collinearity of 3GPP and the proposed model at 2.4 GHz ($d_{MS}$ represents the distance between two Rxs, the simulated scenario is indoor office NLOS scenario in\cite{b1}).}
   \label{cmc}
\end{figure}

% First,  an example of trajectories of the receiver and clusters was given to visualize the cluster birth-death process. Then the statistical properties including ACF, CCF and PDP of 3GPP TR38.901 channel model, QuaDRiGa channel model and the proposed model were compared. In addition, the introduction of Measurement data of ACF and CCF helps to validate the proposed model. At last, we ran multi-link and multi-frequency simulations to see how 3GPP channel model and the proposed model behave under different frequency bands and links.\\
\indent Fig. \ref{trajectory} displays the trajectories of the Rx and clusters, with arrows denoting the moving directions. The objects can be either static or have linear and circular trajectories. The cluster disappearance is shown in dashed line and appearance is demonstrated in solid line. Flexible trajectory setting makes the model more realistic and brings about different statistical properties from the other two models.
\\
\indent A comparison is made between PDPs of 3GPP\cite{b1}, QuaDRiGa\cite{b2}, and the proposed models, as is shown in Fig.~\ref{pdp}. Fast fading model in \cite{b1} does not consider the time-variety of delay and power. Drifting process in \cite{b2} ensures that delays and powers slowly drift in time due to the fixed positions of clusters. Involving cluster movement and birth-death process, the variation of delays and powers becomes more conspicuous. In addition, the effect of virtual delays can be distinguished in the serpentine traces of delays derived from moving clusters.
\\
\indent Fig. \ref{ccf_mea} provides the comparison of spatial CCFs of 3GPP\cite{b1}, QuaDRiGa\cite{b2}, the proposed model, and the measurement data in\cite{CCF_mea}. It can be found that the three models can accurately fit with the descending part of measurement data. This is because they are 3D models which consider not only azimuth angles but also elevation angles. Since the proposed model neglects the effect of spherical waves, there is a slight difference of spatial CCFs between the proposed model and QuaDRiGa in the simulated antenna size.\\
\indent Fig. \ref{cmc} compares the CDF of CMC of the proposed model and that of 3GPP\cite{b1}. In the proposed model, the value of CMC obviously decreases with the increase of the receiver distance, which shows the correlation between links is decreasing. However, the CMC value in\cite{b1} is not sensitive to the change of receiver distance and smaller than the proposed model. Beacause only  the correlation of LSPs between links are considered.
\section{Conclusions}
A novel 3D non-stationary multi-frequency multi-link wideband MIMO channel model has been proposed in this paper, in which time evolution is featured by cluster birth-death process and correlations between different frequency bands and links are taken into account. The CCFs of 3GPP, QuaDRiGa, and our model are compared with measurement data. It turns out that the three models can fit the measurement data well, since elevation angles are considered. In addition, our model exhibits more flexibility in terms of describing time-variant channels with moving cluster. Moreover, the link correlation has been well modeled by using SoS method to generate LSPs and SSPs. \\
\indent In the future, we will compare other important statistical properties of the three models and futhur analyze multi-frequency correlation of the proposed model. Besides, we will carry out some channel measurements to validate the investigated properties of the proposed model. 
 \section*{Acknowledgment}
\small {This work was supported by the National Key R\&D Program of China under Grant 2018YFB1801101, the National Natural Science Foundation of China (NSFC) under Grant 61960206006 and Grant 61901109, the Frontiers Science Center for Mobile Information Communication and Security, the High Level Innovation and Entrepreneurial Research Team Program in Jiangsu, the High Level Innovation and Entrepreneurial Talent Introduction Program in Jiangsu, the Research Fund of National Mobile Communications Research Laboratory, Southeast University, under Grant 2020B01, the Fundamental Research Funds for the Central Universities under Grant 2242019R30001, the Huawei Cooperation Project, and the EU H2020 RISE TESTBED2 project under Grant 872172, the National Postdoctoral Program for Innovative Talents under Grant BX20180062.}

\end{document}